\documentclass[a4paper,twocolumn,floatfix,longbibliography]{revtex4-1}
\usepackage{graphicx}
\usepackage[utf8]{inputenc}
\usepackage{siunitx}
\usepackage{pifont}
\usepackage{MnSymbol}
\usepackage{amsmath}
\usepackage[hidelinks]{hyperref}
\usepackage{url}

\usepackage{breakurl}
\hypersetup{breaklinks=true}
\usepackage[english]{babel}

\begin{document}

\title{A low-cost and reliable laser shutter interlock using a software-command interface}

\author{Joshua P. Rogers, Andrew J. Murray}

\affiliation{Photon Science Institute, School of Physics \& Astronomy, University of Manchester, Manchester M13 9PL, United Kingdom}

\date{\today}

\begin{abstract}
    A simple and low-cost laser interlock is presented that operates via software commands issued by an ESP32 microcontroller.  The architecture of the device is constructed to ensure the laser output is shut off in the event of either an open circuit on the interlock signal line from the laser enclosure or loss of power to the device.  Unintentional exposure to the laser beam is prevented by overruling local controls (such as a keypad), until both the enclosure is re-interlocked and the user actively intervenes.  The device presented is designed to close the mechanical shutter of a Spectra-Physics Millennia Pro pump laser while it continues to operate internally.  The hardware and coding are versatile enough to be deployed on any instrument that receives software commands via a serial interface.
\end{abstract}


\maketitle

\section{Introduction}

Laser safety in a working research laboratory is extremely important.  In the UK in recent years, many laser users have had to suspend activities for weeks or months to make improvements in laser safety measures following inspections by the Health and Safety Executive.  In the hierarchy of controls for the safety of workers~\cite{Chase2019}, engineered containment of a hazard is preferred over both administrative controls and personal protective equipment (PPE). Containment is superseded only by elimination or substitution of the hazard.  Many laser laboratories use high power class 3B and 4 lasers, which exceed maximum permissible exposure irradiance limits to the eye and often also to the skin.  If it cannot be eliminated, the laser emission hazard therefore requires robust containment infrastructure to protect all laboratory occupants and reduce reliance on PPE, such as laser goggles and gloves.  Ideally, all laser beams are always fully enclosed and inaccessible to all lab users.  However, a door or curtain interlock is an essential component in these safety measures, particularly when the room itself becomes the laser enclosure while open-beam work is in progress. This situation occurs during manual alignment of the optical path, or when adjusting a laser cavity to optimize its output.  

The ideal interlock is `fail-safe', such that not only a discontinuity in the enclosure but also power or component failure in the interlock itself will immediately halt laser emission.  Interlocks therefore typically employ a switching circuit that is closed only when the enclosure is shut.  Depending on the type of laser system that is used, the interlock may directly shut off power to the source of emission or it may mechanically obstruct the beam using a shutter.  Whereas cutting off the power to the laser will definitively remove the hazard, consideration must also be given to whether this is practicable under research conditions.  This is particularly true if the laser or downstream components in the optical chain could be damaged by sudden loss of power. Strategically positioned mechanical shutters are therefore often more appropriate in these situations as they are less disruptive to research activity while providing the same degree of protection to users.

The interlock system described here has been designed to interface with the 15~\si{\watt} Spectra-Physics Millennia Pro laser, which is used to pump several tuneable laser systems in our laboratory. This laser features an electrical interlock interface in series with its power supply, which ensures that emission stops immediately when the laser enclosure interlock circuit is opened. Once the enclosure interlock switch is re-engaged the laser must then be restarted and the warming-up procedure to reach full power and stability can be implemented. This can introduce significant delay and disruption to research activity, particularly when frequent access to the laser enclosure is required. Despite being highly dependable, this is not a practicable method for interlocking the laser system to its enclosure.  Although interlock overrides are often used to circumvent this problem, disabling a safety system, even temporarily, increases the probability of accidental exposure to the hazard.

The Millennia also features a gravity-fed mechanical shutter inside the laser head, operated by a handheld controller.  The shutter is normally-closed and, rather than interrupting power to the laser head, merely physically obstructs the beam, keeping it within the Millenia's housing.  The shutter can also be actuated by software commands via an RS232 interface on the rear of the power supply. Since shutting the shutter ensures complete containment of the beam while maintaining stable laser operating conditions internally, it is clearly preferable to use this method to eliminate the hazard when the enclosure is opened during open-beam work. 

In this design note a simple and low-cost interlock solution is presented that satisfies the requirements of being fail-safe, reliable and practicable. This solution utilises an ESP32 microprocessor on a commercial development board that interfaces with the Millennia's software command interface to actuate its mechanical shutter. The ESP32 continuously monitors the status of both the interlock and the shutter and will shut off the laser beam if the enclosure is opened. The laser power supply therefore remains operational throughout so that the long term stability of the laser is not disturbed.

An interlock that utilises software commands can have more potential points of failure compared to a simple hardware interlock. It is therefore important to ensure that it is fail-safe under all operating conditions.  The ESP32 microcontroller serves this purpose well because it is a robust device that operates independently of an operating system and does not need to interface to a network. The code that has been written for the microcontroller only performs the dedicated task of monitoring the interlock status and writing commands to the RS232 interface. This ensures continuous, reliable and safe operation once installed and does not require the user to override its function.

\section{Methods}

\begin{figure}
    \centering
    \includegraphics[width=85mm]{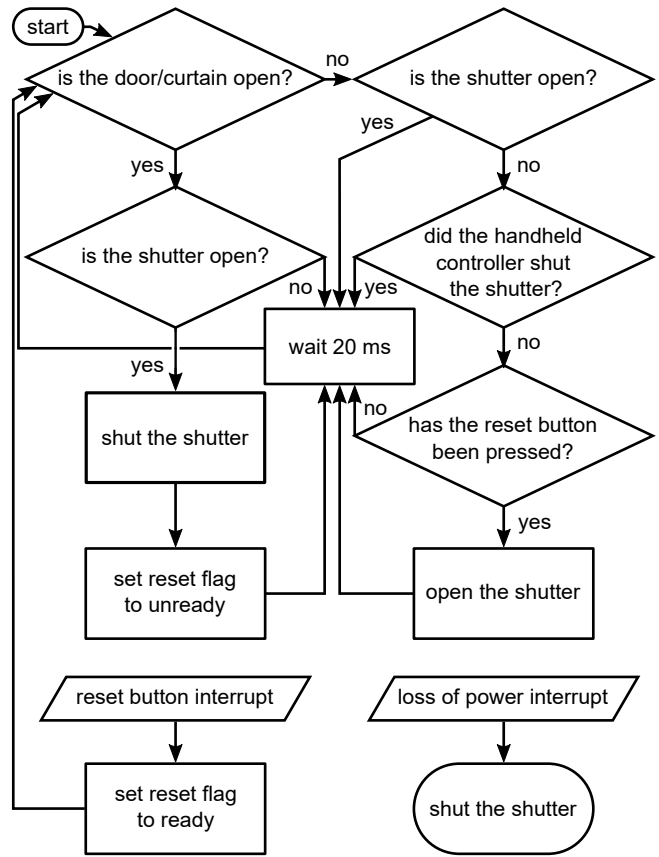}
    \caption{A flow diagram showing the logic of the state machine coded on the ESP32.}
    \label{Fig:Logic}
\end{figure}

The principle of the interlock is a basic state machine, the logic of which is shown in Fig.~\ref{Fig:Logic}.  If the enclosure interlock switch opens, a serial command is sent to the laser to immediately shut the mechanical shutter if it is open.  When the enclosure interlock is re-engaged, a physical reset button on the interlock box must be pressed before the laser shutter will open again. This ensures that the user is active in controlling the safety aspects of the system.  A 20~\si{\milli\second} loop that continuously checks the status of the enclosure interlock was chosen over interrupt logic due to severe bouncing in the interlock switch signal.  This loop duration is long enough to prevent rapid repeated actuation of the shutter due to this switch bounce.  It is also short enough to act on the same time scale as the motion of the shutter and appear instantaneous to the user; including the transmission of the serial data, the shutter will always begin to close within 32~\si{\milli\second} of the enclosure interlock switch opening.

\begin{figure}
    \centering
    \includegraphics[width=85mm]{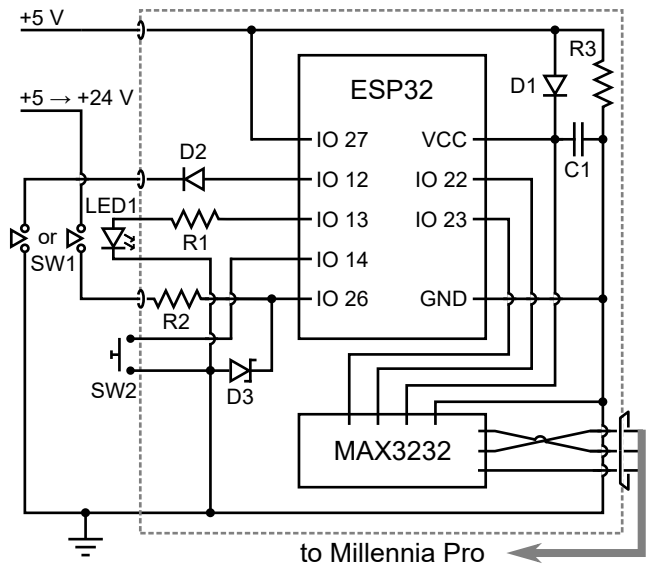}
    \caption{The electrical schematic of the interlock.  The dotted grey boundary represents the box containing the ESP32, MAX3232 IC, bespoke PCB and panel connectors.  Both varieties of enclosure interlock circuit (passive and active) are shown with their respective switches on the left.  The thick grey arrow on the bottom right represents the RS232 cable connected to the Millennia Pro power supply.  The values for the components are listed in Table~\ref{Tab:Components}.}
    \label{Fig:Schematic}
\end{figure}

\begin{figure}
    \centering
    \includegraphics[width=85mm]{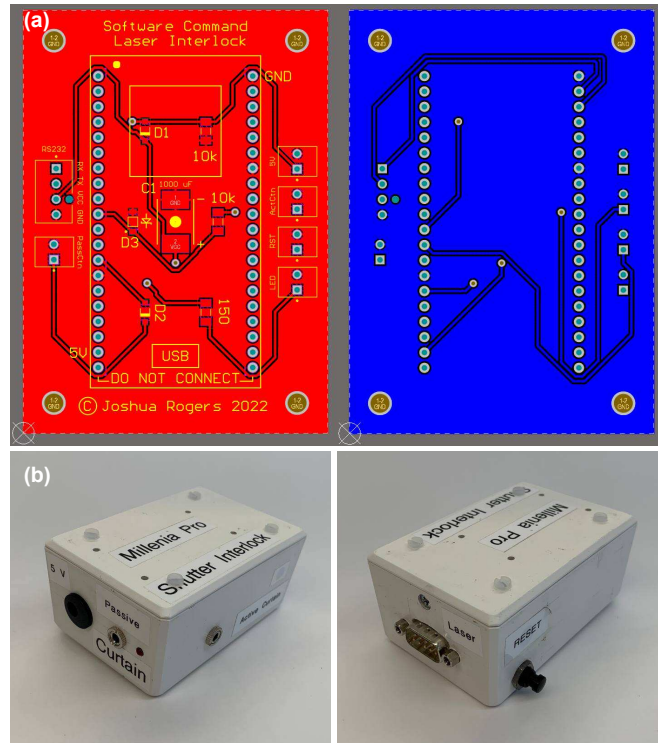}
    \caption{(a) The bespoke PCB front (red) and back (blue) planes. (b) Pictures of the finished interlock box, measuring 9~$\times$~6~$\times$~4~\si{\centi\meter}.}
    \label{Fig:PCB}
\end{figure}

The ESP32 development board that controls the interlock sits in a small box and is mounted on a bespoke PCB that delivers +5~\si{\volt} DC power and breakout connections. It is wired to both the enclosure interlock switch and the laser power supply RS232 interface.  A schematic of the interlock circuit is shown in Fig.~\ref{Fig:Schematic} and the values for the components are listed in Table~\ref{Tab:Components}.  The ESP32 monitors the status of the enclosure interlock via two of its general-purpose input/output (GPIO) pins.  For a `passive' normally-open circuit, switch SW1 shorts to ground when the enclosure interlock is shut. The GPIO pin 12 monitors this switch and is set to use an internally enabled pull-up resistor, so that the input is high when the curtain is open and is low when the switch SW1 is closed.  A 1N4148 diode (D2) protects this input from any accidental positive voltage input up to 100~\si{\volt}.  If an `active' enclosure interlock system is used (with a range from +5VDC to +24VDC), then this is input via a separate socket (see Fig.~\ref{Fig:PCB}). This active interlock voltage is monitored by GPIO pin 26, which is protected by a 4.7~\si{\volt} Zener diode (D3) as shown. The 10~\si{\kilo\ohm} resistor R3 limits the current from the interlock supply while ensuring the Zener diode is switched on. 

Pin 13 is configured as an output and provides a signal to an indicator LED (LED1) on the side of the box that illuminates when the shutter is open.

Pin 27 monitors the ESP32 power supply.  If it is pulled low (eg by a +5~\si{\volt} DC power supply failure), the 1~\si{\milli\farad} capacitor (C1) across the power input holds enough charge to ensure that a final command can be sent to close the shutter before the ESP32 stops operating.  This feature ensures that the interlock is fail-safe against accidental loss of power.

Pin 14 is set with the pull-up resistor enabled and is used to interrupt the code loop to indicate when the reset button is pressed.  This interrupt flag ensures that open beam work cannot resume until the user intervenes directly.  The software allows the handheld laser controller to be used normally while the enclosure interlock is shut, however the ESP32 system overrides the manual controller to ensure the shutter remains closed while the enclosure is open.

\begin{table}
    \centering
    \begin{tabular}{c c}
        \hline \hline
        \vspace{-10pt} \\
        component & value \\
        \hline
        \vspace{-9pt} \\
        R1 & 150~\si{\ohm}\\
        R2 & 10~\si{\kilo\ohm} \\
        R3 & 1~\si{\kilo\ohm} \\
        C1 & 1~\si{\milli\farad} \\
        D1 & RR264M-400TR \\
        D2 & 1N4148W-7-F \\
        D3 & MMSZ4688T1G \\
        LED1 & 703-0090 \\
        \hline
        \hline
    \end{tabular}
    \caption{Values for the components in Fig.~\ref{Fig:Schematic}.}
    \label{Tab:Components}
\end{table}

Pins 22 and 23 are assigned respectively to receive and transmit serial data.  The serial interface of the ESP32 operates on normal transistor-transistor logic (TTL) levels. The Millennia communicates over RS232, which uses inverted high and low logic levels in the range $\pm 3 - 15$~\si{\volt}~\cite{shawn_2020}.  The serial communication protocol of the Millennia power supply operates at 9600 baud, with no parity bit, 8 data bits, 1 stop bit and no hardware flow control (config SERIAL\_8N1)~\cite{Millennia2005}.  To translate between the TTL and RS232 protocols, a MAX3232 IC is used as shown in Fig.~\ref{Fig:Schematic}.

It is important to note that there are several ESP32 development boards available that use different pin-outs. The bespoke PCB in this interlock is compatible with the 38-pin NodeMCU ESP32 board.  Its design is shown in Fig.~\ref{Fig:PCB} (a), together with images of the finished interlock box (b).  The ESP32 is mounted onto the PCB on two rows of header sockets, which serve to breakout the GPIO pins to Molex KK 254 series connectors.  This modular design ensures that if the board, the ESP32 or a port on the box fails, replacements can be inserted easily.  The bill of materials (10.48420/20037140), ESP32 software (10.48420/20037176) and PCB design files (10.48420/20037185) are all available online at their respective DOIs.

\section{Discussion}

The interlock system described in this design note was built to ensure versatility, reliability and modularity.  It features two physical input ports for the enclosure interlock switch, depending on whether the external circuit is active or passive. If active, the system can accept any voltage between 5~-~24\si{\volt}.  The ESP32 platform is simple, robust and low maintenance, with only periodic functionality checks being required.  Components are easily swapped in case of failure.  Although the software code is written specifically for the Millennia Pro pump laser, simple edits are all that are required for it to be deployed on any instrument that is controlled by serial software commands.  The interlock system described here costs around £75 per unit, including the bespoke PCB manufactured at a prototyping fabricator.

\section{Conclusions}

Operational safely is the first priority in any laboratory that uses high power laser systems.  Engineered solutions that are reliable, cost-effective and practicable are therefore essential for maximising uptime and project resources.  The interlock system described here can be implemented with only a small degree of prior experience in electronics, or it can even be used as an exercise to develop the necessary skills in this area.  These principles are also easily configurable to enable hardware-triggered automation in any similar laboratory device that uses an RS232 interface.

\section{Acknowledgements}

This work was supported by the Engineering and Physical Sciences Research Council (EPSRC), United Kingdom, through research grant EP/V027689/1.  Thanks are also given to Derrick Bradshaw in the PCB fabrication lab of the University of Manchester Department of Engineering for manufacturing the bespoke boards.

\raggedright
%\Urlmuskip=0mu plus 1mu\relax
\bibliography{interlock}

\end{document}